\documentclass[fleqn,usenatbib]{mnras}

\usepackage{newtxtext,newtxmath}

\usepackage[T1]{fontenc}

\DeclareRobustCommand{\VAN}[3]{#2}
\let\VANthebibliography\thebibliography
\def\thebibliography{\DeclareRobustCommand{\VAN}[3]{##3}\VANthebibliography}

\usepackage{graphicx}	
\usepackage{amsmath}	

\title[2021 October Jupiter impact model]{Modelling the optical energy profile of the 2021 October Jupiter impact flash}

\author[K. Arimatsu et al.]{
Ko Arimatsu,$^{1}$\thanks{E-mail: arimatsu.ko.6x@kyoto-u.ac.jp}
Kohji Tsumura,$^{2}$
Fumihiko Usui$^{3}$
and Jun-ichi Watanabe$^{4}$
\\
$^{1}$The Hakubi Center/Astronomical Observatory, Graduate School of Science, Kyoto University Kitashirakawa-oiwake-cho, Sakyo-ku, Kyoto 606-8502, Japan\\
$^{2}$Department of Natural Science, Faculty of Science and Engineering, Tokyo City University, Setagaya, Tokyo 158-8557, Japan\\
$^{3}$Institute of Space and Astronautical Science (ISAS), Japan Aerospace Exploration Agency (JAXA), 3-1-1 Yoshinodai, Chuo-ku, Sagamihara, Kanagawa 252-5210, Japan\\
$^{4}$National Astronomical Observatory of Japan, 2-21-1 Osawa, Mitaka, Tokyo 181-8588, Japan
}

\date{Accepted XXX. Received YYY; in original form ZZZ}

\pubyear{2023}

\begin{document}
\label{firstpage}
\pagerange{\pageref{firstpage}--\pageref{lastpage}}
\maketitle

\begin{abstract}
We have conducted numerical simulations to reproduce the observed optical energy profile of the 15 October 2021 (UT) impact flash on Jupiter, which was the largest and the most well-observed flash event detected by ground-based movie observations.
The observed long-duration ($\sim 5.5~{\rm s}$) optical emission can be reproduced by an impact of an object with an exceptionally small angle of entry relative to the horizontal.
The apparent lack of the impact debris feature despite the large impact object was possibly due to the shallower angle of entry ($\le 12\degr$), 
which resulted in the lower ablation per unit volume at altitudes higher than $50 \, {\rm km}$, 
and the volume densities of the ablated materials were too low to allow the debris particulates to coagulate.
The absence of temporal methane absorption change in the observed flash spectrum is consistent with the best-fit results. 
The model better fits the observed optical energy profile for weaker material (cometary and stony) cases than for metallic ones.
Based on the simulation results, prospects for future observations of impact flashes are discussed.
\end{abstract}




\section{Introduction}
\label{sec:intro}

Optical flashes resulting from the impact of unidentified interplanetary objects on Jupiter have been serendipitously observed through ground-based amateur observations \citep{Hueso2010-xv, Hueso2013-ki,Hueso2018-nr,Sankar2020-lm}.  
Their emission characteristics should demonstrate the radiative consequences of decameter-sized impacts on planetary atmospheres, which could potentially threaten human society \citep{Jenniskens2019-tg,Boslough1997-pi,Boslough2008-ae} but are unknown due to their infrequent occurrence on Earth \citep{Brown2002-aq}. 
Detailed investigations of these impacts also provide a unique opportunity to explore the abundance and physical characteristics of small objects in the outer solar system, 
as it is impossible to detect them directly \citep{Hueso2013-ki,Hueso2018-nr,Giles2021-sj}.
\citet{Sankar2020-lm} recently developed a fragmentation model for superbolides on Jupiter. 
They found that comparing the model with the observed light curve of the 2019 Jovian flash can constrain the angle of entry and material make-up of the impacting object.
Modellings of the Jovian impact flashes thus can provide information on complex situations of the impacts that should be essential for better understanding the impact objects and their effects on planetary atmospheres.
However, the lack of observational constraints on the spectral information of the flashes hampered from making applications of the modellings without poor approximations to the conversion from the observed brightness to the total optical energy of the flash.

On 15 October 2021, 
a new impact flash on Jupiter was detected by an optical multi-band observation system, Planetary ObservatioN
Camera for Optical Transient Surveys (PONCOTS; \citealt{Arimatsu2022-if}).
Since the PONCOTS system achieved a high-cadence and three-band simultaneous observation of the impact flash for the first time, 
it allows obtaining its physical parameters
without the need for poor approximations on the spectral information.
The mass and diameter of the impact object were estimated to be $\sim 4 \times 10^6 ~ {\rm kg}$ and $\sim 12 \-- 25 ~ {\rm m}$, respectively (\citealt{Arimatsu2022-if}, see also Section~\ref{subsec:chi2}).
The total kinetic energy of the impact object is estimated to be $\sim$ two megatons 
(hereafter Mts)
of TNT, 
an order of magnitude greater than that of previously detected flashes on Jupiter and thus the largest impact flash observed in the solar system since the impacts of the fragments of the comet Shoemaker–Levy 9 (hereafter SL9) on Jupiter \citep{Hammel1995-nn, Crawford1997-sg, Harrington2004-kf}.
We should also note that the impact energy is comparable with the Tunguska impact on Earth in 1908 \citep{Boslough1997-pi,Boslough2008-ae}.
Modelling the 2021 October impact flash 
discovered by PONCOTS
(hereafter "PONCOTS flash") 
thus provides a unique opportunity for understanding the nature of such large flashes that could threaten human society. 
In addition, more detailed analyses of 
the PONCOTS flash
are required to understand its observed characteristics that are unique and possibly 
different from
the previous impact events.
First, 
the duration of the flash was approximately 5.5~sec, which is much longer than those of previously detected flashes ($\sim 1\--2~{\rm sec}$, \citealt{Hueso2013-ki,Hueso2018-nr}).
Furthermore, even though its estimated large mass and long flash duration indicate that the impact object could have reached the lower atmosphere, no evident temporal change in the methane absorption in the flash spectrum was observed (see Section~\ref{subsec:ch4}). 
In addition, despite the enormous estimated mass of the impact object, no apparent impact feature was found after the flash, which could be
different from 
the SL9 impact results \citep{Hammel1995-nn}.
Modelling the emission characteristics of the large impact flash would help to find a possible scenario consistent with its observed features.

This paper presents the modelling of the optical energy profile of 
the PONCOTS flash
on Jupiter.
In Section~\ref{sec:obs}, we present the outline of the 2021 October impact observed by the PONCOTS observation system.
We introduce a fragmentation model and assumed conditions used to fit the light curves in Section~\ref{sec:ana}.
Results of the fit and discussions are presented in Section~\ref{sec:res}.
We summarize the conclusions in Section~\ref{sec:con}.

\section{Overview of the PONCOTS flash}
\label{sec:obs}

The PONCOTS system consists of an $D = 0.279~{\rm m}$ Schmidt-Cassegrain optical tube for amateur astronomers (Celestron C11) equipped with high-cadence monochrome complementary metal-oxide semiconductor (CMOS) cameras (QHY5III-290M camera with a SONY IMX 290 sensor for camera modules of the two shorter-wavelength beams split by two dichroic mirrors and the Planetary one Neptune-CII camera with a SONY IMX464 sensor for the longest-wavelength beam).
In the 2021 observation campaign, we used two of the three wavelength bands, 
the $V$ ($\lambda = 505\--650~{\rm nm}$) and ${\rm CH_4}$ ($880\--900~{\rm nm}$).
In addition, we adopted an artefact image in the $V$ band data as another wavelength band image named "Gh-band" ($680\--840~{\rm nm}$, see \citealt{Arimatsu2022-if}).
The flash was detected in all three bands.

Until now, we have received three observation reports (two in Japan and another in Singapore) by amateur astronomers about the same impact flash.
These observation reports indicate that the present flash unambiguously occurred on Jupiter, not in the terrestrial atmosphere. 
We should note that two of the three amateur observations recorded movie data of the flash, 
 and the recorded data contain saturated pixels of the flash location.

Signals obtained by aperture photometry of the flash in the PONCOTS three-band images were calibrated with a spectrophotometric standard star (HR 7950; V = 3.78 mag; Spectral type A1V; \cite{Hamuy1992-nw}, see \citet{Arimatsu2022-if} for details). 
Fluxes for individual frames were binned into 0.5~s time bins to provide temporal variations of the spectrum with sufficient signal-to-noise ratios.
Since strong backward reflection of the Jovian upper clouds significantly contributed to the observed flash fluxes, 
we estimated and corrected the contribution of the cloud-reflection component based on the wavelength-dependent scattering phase functions of the Jovian surface provided by \citet{Heng2021-my}.
The cloud-reflection component is approximately 70\%, 60\%, and 30\% of the observed fluxes in the V, Gh, and ${\rm CH_4}$ bands, respectively. 
Details of the procedure will be given in a separate paper.

After the cloud-reflection correction, we then fitted each 0.5 s bin SED with a single-temperature blackbody radiation spectral model to derive the optical energy and the effective temperature. 
As already shown in \citet{Arimatsu2022-if}, 
SEDs for most bins were approximated by a single-temperature blackbody spectrum with the best-fit temperature being $8300 \pm 600 $ K without evident temporal variation. 
The best-fit optical energy for each time bin is shown in Figure~\ref{fig:fit}.
Total optical energy $E_0$ was determined as the sum of the optical energy for individual bins,
$E_0 = 1.8^{+0.9}_{-0.2}\times 10^{15} \, {\rm J}.$
Total kinetic energy $E_T$ was derived from $E_0$ through the relationship adopted from \citet{Brown2002-aq},
\begin{eqnarray}
E_T & = & \eta^{-1} \, E_0 \\
\eta & = & 0.12 \, E_0^{0.115}
\end{eqnarray}
where $\eta$ is the optical energy efficiency, and $E_T$ and $E_0$ are in kiloton TNT (kt; $1\, {\rm kt} = 4.185 \times 10^{12} \,  {\rm J}$).
$E_T$ was determined to be $E_T={7.4}^{+3.3}_{-0.9} \times{10}^{15}\, \mathrm{J}$ with $\eta = 0.24$.

\section{Modelling of the PONCOTS flash}
\label{sec:ana}

\subsection{Ablation and fragmentation model}
\label{sec:model}

We compared the optical energy profile derived from the previous study (\citealt{Arimatsu2022-if}, see also Section~\ref{sec:obs} and Figure~\ref{fig:fit}) with an ablation and fragmentation model based on a code provided by \citet{Sankar2020-lm}, 
which was developed from energy deposition models for terrestrial superbolides \citep{Avramenko2014-nw,Wheeler2017-jk}.
The velocity $v$, mass $M$, height $h$, and flight angle $\theta$ are given by the following differential equations:
\begin{eqnarray}
\frac{{\mathrm d}v}{{\mathrm d}t} &=& -\frac{C_D\, S\rho_a\left(h\right)\, v^2}{2M}+g \sin{\theta}, \\
\frac{{\mathrm d}M}{{\mathrm d}t}&=&-\frac{S\, \sigma_{\mathrm{ab}}\, \rho_a\, \left(h\right)\, v^3}{2}, \\
\frac{{\mathrm d}h}{{\mathrm d}t} &=& -v \sin{\theta}, \\
\frac{{\mathrm d}\theta}{{\mathrm d}t} &=& \frac{g \cos{\theta}}{v}+\frac{v \cos{\theta}}{R_J+h},
\label{eq1}
\end{eqnarray}

where $C_D$ is the drag coefficient with $C_D\ =\ 0.92$, 
following the previous studies \citep{Carter2009-nb}, 
$S$ is the cross-section area, 
$\sigma_{\mathrm{ab}}$ is the ablation coefficient (the amount of evaporated material per unit energy), 
$R_J$ is the Jovian radius with $R_J=7\times{10}^4\, \mathrm{km}$, 
and $g$ is the gravitational acceleration with $g=25\, {\rm m\, s^{-1}}$, respectively. 
$\rho_a$ is the atmospheric density as a function of height $h$.
In the present study, 
$\rho_a$ is calculated using the vertical 
temperature and density profile of Jovian atmosphere from \citet{Moses2005}.

In this model, fragmentation begins when the ram pressure $\rho_av^2$ exceeds the bulk strength of the object. At this point, an object is fragmented into the number $N_\mathrm{f}$ of equal-sized objects given  with a bulk density of the object $\rho$ by 
\begin{equation}
N_{\mathrm{f}}=\frac{16 S^3  \rho}{9\pi M^2}.
\end{equation}
The mass of each fragment $M_\mathrm{f}$ is thus given by
\begin{equation}
    M_\mathrm{f}=\frac{M}{N_\mathrm{f}}=\frac{9\pi M^3}{16S^3 \rho}.
\end{equation}
Since the fragmentation break is thought to eliminate larger structural weaknesses, 
the smaller fragments are assumed to be stronger than the original body. 
The bulk strength of the smaller fragment 
is thus assumed to be given by the following Weibull-like exponential scaling relation 
\begin{equation}
    \sigma_\mathrm{f}=\sigma_0\left(\frac{M_\mathrm{f}}{M_0}\right)^{-\alpha},
\end{equation}
where $\sigma_\mathrm{f}$ and $M_\mathrm{f}$ are the strength and mass of the fragment and $\sigma_0$ and $M_0$ are the initial strength and mass. 
The strength scaling parameter $\alpha$ is set to be a free parameter for the fit.
With the fragmentation effect, the cross-section area is thus given by the following equations
\begin{equation}
    \frac{{\mathrm d}S}{{\mathrm d}t} = 
\begin{cases}
\frac{2}{3}\frac{S}{M}\frac{{\mathrm d}M}{{\mathrm d}t} & \rho_av^2 < \sigma_\mathrm{f},\\
\frac{2}{3}\frac{S}{M}\frac{{\mathrm d}M}{{\mathrm d}t}+C_\mathrm{f}\frac{S\sqrt{\rho_av^2-\sigma_\mathrm{f}}}{M^{1/3}\rho^{1/6}} & \rho_a v^2>\sigma_\mathrm{f},
\end{cases}
\end{equation}
where $C_\mathrm{f}$ is a dimensionless free parameter that accounts for the dispersive effects of these fragments.

\subsection{Test conditions and fitting procedures}
\label{subsec:chi2}

\begin{table}
	\centering
	\caption{Input prameters for the different material cases.}
	\label{tab:input_param}
	\begin{tabular}{lccc} 
		\hline
		Case        & $\rho$                   & $\sigma_0$  & $\sigma_{\rm ab}$ \\
		            & (kg ${\mathrm m^{-3}}$)  &   (Pa)      & (kg ${\mathrm J^{-1}}$)\\  
		\hline
		cometary    & 500           & $1 \times 10^4$  & $2 \times 10^{-8}$\\
		stony      & 2500          & $5 \times 10^5$  & $2 \times 10^{-9}$\\
		metallic    & 5000          & $10^7$           & $10^{-8}$\\
	\end{tabular}
\end{table}

Since the impact velocity $v_0$ for Jovian impact objects is thought to be comparable with the escape velocity of Jupiter \citep{Harrington2004-kf},
$v_0$ is set to be $v_0 \simeq 60~{\rm km \, s^{-1}}$. 
The mass of the impact object $M_0$ was therefore estimated from $E_T$ to be 
$M_0 = 2~E_T / v_0^2= {4.1} \times 10^6 \, {\rm kg}$. 

Based on the previous studies by \citet{Sankar2020-lm}, we assume three different material cases named
"cometary", "stony", and "metallic".
Input parameters (the bulk density $\rho$, 
the initial strength $\sigma_0$, and 
the ablation coefficient $\sigma_{\rm ab}$)
for these three cases are listed in Table~\ref{tab:input_param}.
The diameter $D$ of the impact object corresponds to $25$, $15$, and $12$ m for the cometary ($\rho = 500$ kg ${\mathrm m^{-3}}$), 
stony ($\rho = 2500$ kg ${\mathrm m^{-3}}$),
and metallic ($\rho = 5000$ kg ${\mathrm m^{-3}}$) cases, respectively, under the assumption of its spherical shape. 
In the previous studies, a wide range of initial strengths $\sigma_0$ was assumed for the metallic case ($\sigma_0 = 2 \times 10^6 \-- 10^8~{\rm Pa}$, \citealt{Chyba1993-ph, Sankar2020-lm}). 
In general, the initial bulk strengths are expected to be lower than the material strengths (up to $10^8~{\rm Pa}$, \citealt{Chyba1993-ph}) due to internal cracking.
However, if we assume $\sigma_0 < 5 \times 10^6~{\rm Pa}$, the resulting synthetic profiles cannot be approximated by the observed profile, because the early flare strengths become too strong due to intense fragmentation and conflict with the early phase of the observed light curve.
We therefore assume $\sigma_0 = 10^7~{\rm Pa}$ for the metallic case.

The synthetic energy profile is fitted to the observed data points by minimizing $\chi^2$;
\begin{eqnarray}
    \chi^2 &=&  \sum_i \frac{\big(L_i - \int_{t_{i {\rm min}}}^{t_{i {\rm max}}}   \eta\frac{{\mathrm d}E }{{\mathrm d}t} ~ {\mathrm d}t\big)^2}{\sigma^2_i} \\
    &\simeq & \sum_i \frac{\big(L_i - \int_{t_{i {\rm min}}}^{t_{i {\rm max}}} \eta \frac{v^2}{2} \frac{{\mathrm d}M }{{\mathrm d}t} ~ {\mathrm d}t\big)^2}{\sigma^2_i},
\end{eqnarray}
where $L_i$ and $\sigma_i$ are the optical energy and its $1\sigma$ error derived from the observation, 
$E$ is the kinetic energy of the model impact object, 
$t_{i {\rm min}}$ and $t_{i {\rm max}}$ are the beginning and the ending time of the data point $i$, respectively.
The best-fit parameters are obtained by an exhaustive search with ranges of $0\degr < \theta \le 90\degr$, $0 \le \alpha \le 1$, and $10^{-1} \le C_f \le 10^1$. The parameter ranges of $\alpha$ and $C_f$ are determined based on discussions of previous bolide modelling studies by \citet{Svetsov1995-cj} and \citet{Hills1993-dw}, respectively.
The error bars in the parameters are determined by ranges that allow an increase of $\chi^2$ by 1 from the minimum value. 
In our assumed conditions, the time derivative term of the velocity can be negligible.
For the optical energy efficiency $\eta$, we use $\eta = 0.24$ (see Section~\ref{sec:obs}).

\section{Results and discussions}
\label{sec:res}

\subsection{The best-fit models for the observed energy profile}
\begin{figure}
	\includegraphics[width=\columnwidth]{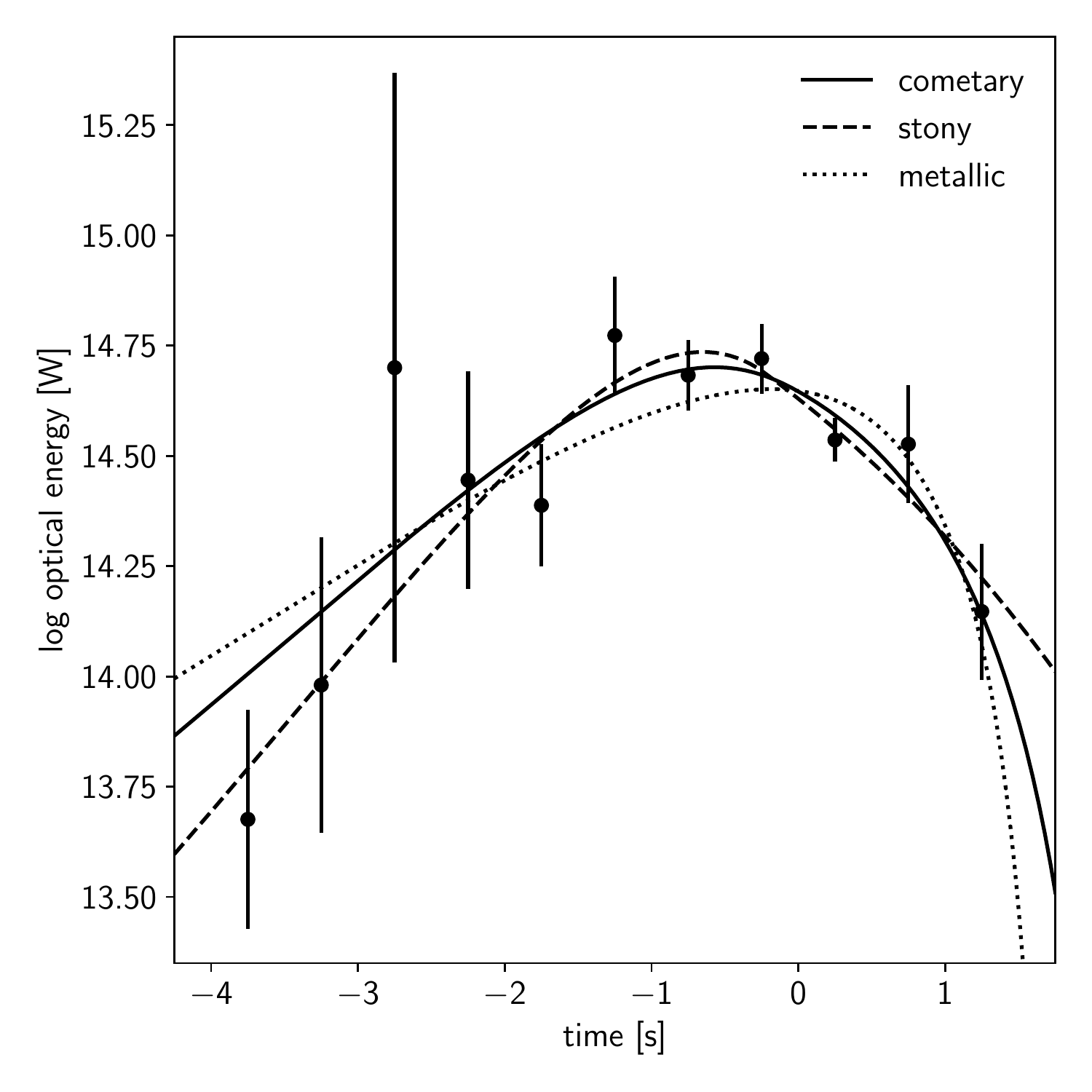}
    \caption{Model-fitting results overlaid with the optical energy profile for each 0.5~s time bin obtained from the PONCOTS observation \citep{Arimatsu2022-if}. 
    The solid, dashed, and dotted curves represent the profile for the cometary, stony, and metallic cases, respectively.}
    \label{fig:fit}
\end{figure}

Figure~\ref{fig:fit} shows the best-fit results of the optical energy profile overlaid with the observed profile.
The observed long duration and the peak intensity can be roughly approximated in all three cases.
For the metallic case, the duration of the rising phase is longer than the observed profile. 
The best-fit parameters and $\chi^2$ values are presented in Table~\ref{tab:fit_result1}.
In either case, 
the best-fit angles of entry are much smaller than that of the SL9 fragments' impacts ($\theta \simeq 45\degr$, \citealt{Crawford1997-sg}) 
or of the 2009 impact ($\theta \simeq 20\degr$, \citealt{Sanchez-Lavega2010-rj}).
We found that smaller angles of entry $\theta$ causing the impact object to ablate longer before reaching the altitude of disruption are required for all the material cases to reproduce the long duration of the flash.
The stony and metallic cases are insensitive to $C_{\rm f}$, 
which is poorly constrained by the present fit
since contributions of fragments are insignificant to the entire cross-section areas for these cases.
Also, $\alpha$ is poorly constrained for the metallic case.
In order to obtain further constraints of the variability of these parameters for the metallic case, we performed the profile fitting with different fixed parameters ($v0$, $\rho$ and $\sigma_0$) in the ranges that can produce profiles comparable to that observed.
However, we found no apparent relationship between the variability and the values of the fixed constants.
The $\chi^2$ values (with 7 degrees of freedom) for the cometary and stony cases are smaller than the metallic case.
The possible weaker material nature is consistent with the assumption that Jupiter-family comets are thought to be a primary source of Jovian impact objects \citep{Levison2000-ze}.

\begin{table}
	\centering
	\caption{Best-fit parameters and $\chi^2$ values for the three material cases.}
	\label{tab:fit_result1}
	\begin{tabular}{lcccc} 
		\hline
		Case        & $\theta$ & $\alpha$ & $C_{\mathrm f}$        & $\chi^2$ \\
		            & (degree) &          &                        &          \\
		\hline 
		cometary    & $6.9^{+3.0}_{-2.2}$  & $0.27^{+0.18}_{-0.12}$  & $1.3^{+1.7}_{-0.3}$ & 6.7 \\
		stony      & $2.8^{+3.6}_{-0.8}$ & $0.06^{+0.12}_{-0.02}$  & $2.0^{+8.0}_{-1.5}$ & 4.5\\
		metallic    & $11.4^{+1.1}_{-0.9}$ & $0.52^{+0.48}_{-0.30}$  & $10.0^{+0.0}_{-9.9}$   & 10.8 \\
	\end{tabular}
\end{table}

\begin{figure}
	\includegraphics[width=\columnwidth]{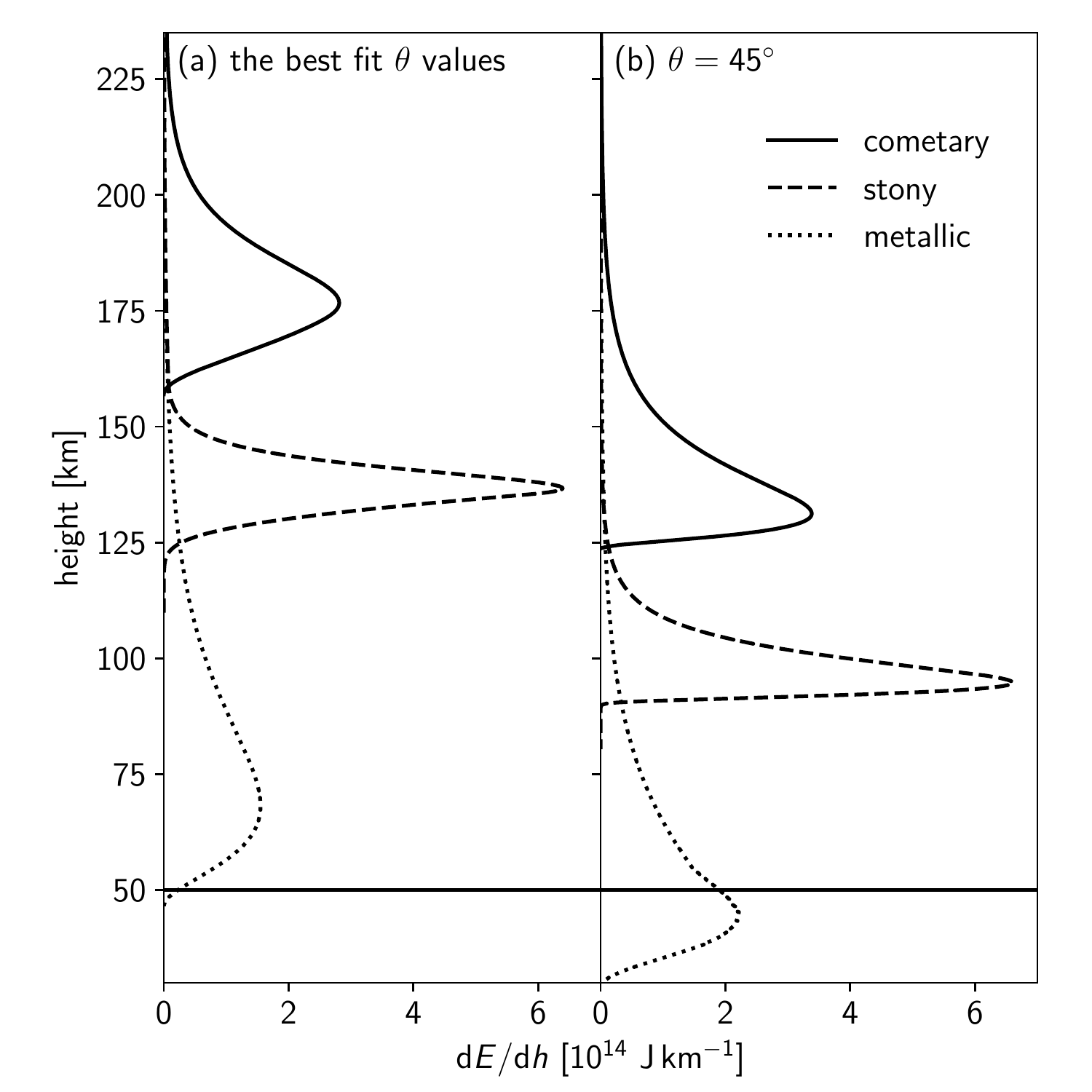}
    \caption{(a) kinetic energy release profiles derived from the best-fit simulation results as a function of the height $h$.
    Solid, dashed, and dotted curves represent the profile for the cometary, stony, and metallic cases.
    (b) the same as (a), but for the models with $\theta = 45\degr$, which is comparable to the angle of entry of the SL9 fragments \citep{Crawford1997-sg}.
    A horizontal line represents the approximate height of the tropopause ($50~{\rm km}$). 
    }
    \label{fig:dedh}
\end{figure}

Figure~\ref{fig:dedh}(a) shows the kinetic energy release profiles derived from the best-fit simulation results as a function of the height $h$, and Table~\ref{tab:fit_result2} presents the peak and end heights and pressures of the impact object for the three material cases.
In the present study, the end height is defined to be the final height where the mass of the remaining fragments becomes smaller than 0.1\% of the initial mass.
For the cometary and stony cases, 
the peak and the end height are much higher than the tropopause (height $\sim 50~{\rm km}$), 
indicating the impact object was ablated and disrupted in the upper and middle stratosphere.
Even for the metallic case with the strongest bulk strength, 
the simulation indicates that the ablation intensity reached its maximum above the troposphere.
For comparison, figure~\ref{fig:dedh}(b) shows the profiles for the models with $\theta = 45\degr$, which is comparable to the angle of entry of the SL9 fragments \citep{Crawford1997-sg}.
The kinetic energy is released at lower heights for the larger $\theta$ cases.
Especially for the metallic case, the peak height is expected to be lower than the tropopause.

\subsection{Non-detection of temporal changes in methane absorption}
\label{subsec:ch4}
\begin{figure}
	\includegraphics[width=\columnwidth]{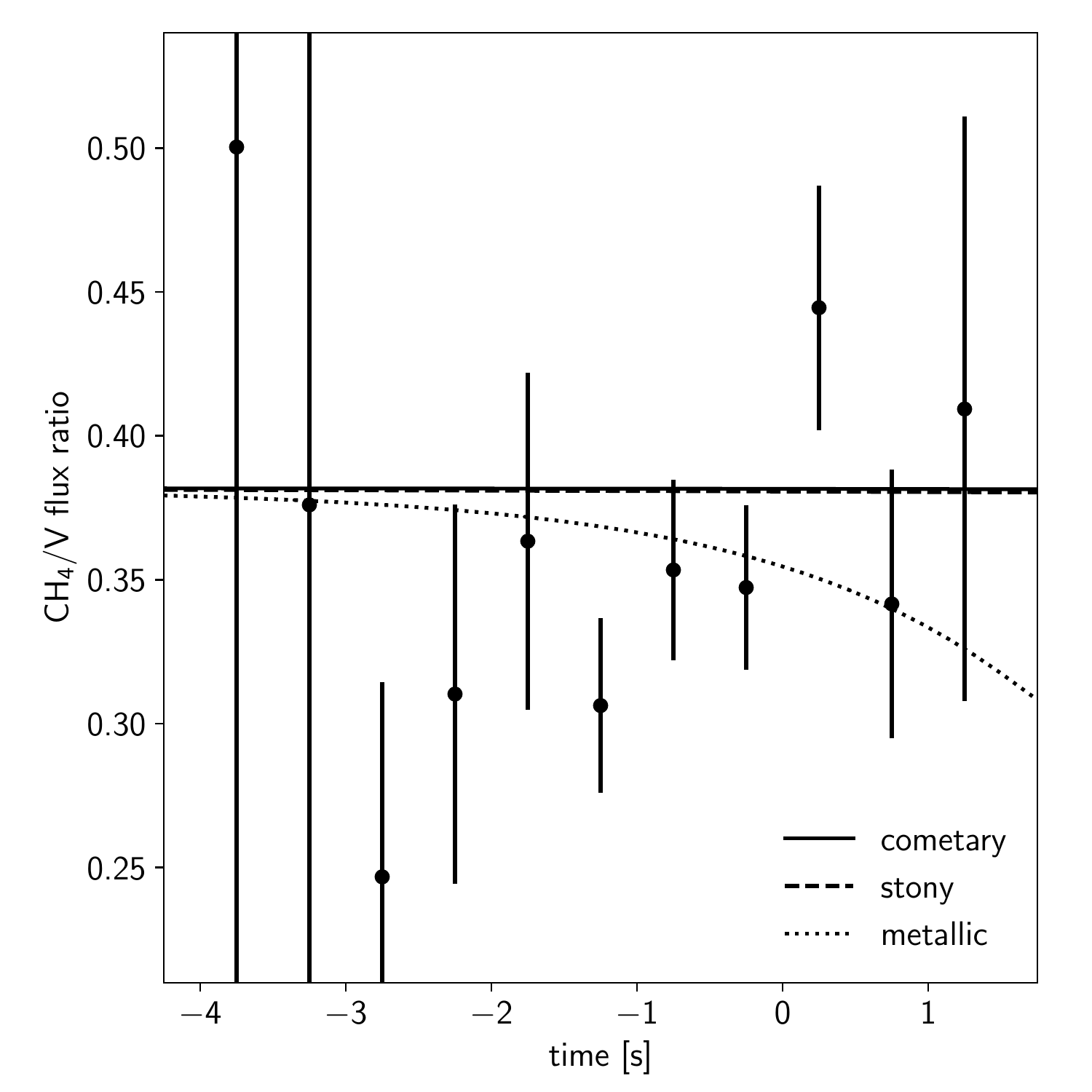}
    \caption{
    Observed ${\rm CH_4}/V$ band flux ratio $f_{{\rm CH_4}/V}(t)$ for the 0.5-s bins overlaid with those derived from the best-fit flash models. 
    Solid, dashed, and dotted curves represent the model flux ratios for the cometary, stony, and metallic cases, respectively. 
    Note that the cometary and stony models almost overlap in this figure.
    The unabsorbed spectrum of the flash is assumed to be a single blackbody with an effective temperature of $8300$ K.}
    \label{fig:col}
\end{figure}

The observed flash radiation from the impact object can be absorbed by methane molecules in the Jovian atmosphere along the line of sight at the ${\rm CH4}$ band, whose central wavelength ($\lambda \simeq 890 \, {\rm nm}$) corresponds to the strong methane absorption band.
The degree of absorption can become stronger 
as the height of the impact object $h$ decreases.
In case of an impact flash caused by a large impact object that could penetrate into the Jovian troposphere, 
a rapid decrease of the ${\rm CH_4}$ band flux caused by strong methane absorption would be expected.
However, as shown in Figure~\ref{fig:col}, we found no clear temporal variations (at least a gradual decrease) of the ${\rm CH_4}$ band flux of the PONCOTS flash relative to its $V$ band flux, $f_{{\rm CH_4}/V}$.
We compare the observed flux ratios with those produced by the best-fit simulation results.
The optical depth at wavelength $\lambda$ along the line of sight to the flash at time $t$, $\tau(t, \lambda)$, is given by
\begin{equation}
\tau(t, \lambda) =\frac{1}{\cos(\mu)}\int^{\infty}_{h(t)} \kappa(P(z), T(z), \lambda ) \, P_{\rm CH4}(z) \, {\rm d}z, 
\end{equation}
where $\mu$ is the angle between the line of sight to the observer and the zenith of the Jovian impact site ($\mu = 26\degr$ for the PONCOTS flash),
$\kappa(P(z), T(z), \lambda )$ is the methane absorption coefficient for pressure $P(z)$ and temperature $T(z)$ at height $z$,
and $P_{\rm CH4}(z)$ is the partial pressure of methane.
$h(t)$ represents the height of the impact object, which is derived from the best-fit optical energy profile model for each material case at time $t$.
To calculate the pressure and temperature dependent $\kappa(P(z), T(z), \lambda )$, we used the methane absorption spectral models developed by \citet{Karkoschka2010-vn}.
We adopted the atmospheric profiles $P(z), T(z),$ and $P_{\rm CH4}(z)$ provided by \citet{Moses2005}.
The model ${\rm CH_4}/V$ band flux ratio $f_{{\rm CH_4}/V}(t)$ at time $t$ is
\begin{equation}
f_{{\rm CH_4}/V}(t) = \frac{F_{\rm CH_4}(t)}{F_V(t)}
\end{equation}
with
\begin{equation}
F_{i}(t) = \frac{\int^\infty_0 R_i(\lambda) \, \exp(-\tau(t, \lambda)) \,  F_{\rm flash}(\lambda) \, {\rm d}\lambda}{\int^\infty_0 R_i(\lambda) \, {\rm d}\lambda} \ \ \ \ {\rm for} \ \ i = {\rm CH_4},V,
\end{equation}
where $R_i(\lambda)$ is the system response of the PONCOTS band $i$ provided by \citet{Arimatsu2022-if},
and $F_{\rm flash}(\lambda)$ is the SED of the flash.
In the present study, $F_{\rm flash}(\lambda)$ is assumed to be a single-temperature blackbody spectrum with a temperature of 8300 K (\citealt{Arimatsu2022-if}, see Section~\ref{sec:obs}).
The $f_{{\rm CH_4}/V}(t)$ for the three material cases are compared in Figure~\ref{fig:col}.
Since the heights for all three material cases are higher than the troposphere, 
the degree of methane absorption at the ${\rm CH_4}$ band is expected to be smaller than $20\%$.
All three cases are thus consistent with the lack of a clear temporal change of the flux ratio.

\begin{table}
	\centering
	\caption{Results of the peak and end locations for the three material cases.}
	\label{tab:fit_result2}
	\begin{tabular}{lcccc} 
		\hline
		Case        & Peak height & Peak pressure  & End height & End pressure \\
		            & (km)        &   (hPa)  & (km)        &   (hPa)         \\
		\hline 
		cometary    & 176.7  &   0.27  & 157.0  & 0.60\\
		stony      & 136.8  &   1.4  &  119.6 & 2.8 \\
		metallic    & 68.4   &   27 &  46.4  & 85 \\
	\end{tabular}
\end{table}

\subsection{Non-detection of debris features}
As noted in \citet{Arimatsu2022-if}, 
the PONCOTS ${\rm CH_4}$ band images obtained 16 minutes after the impact did not show impact-debris features at the site.
Later in situ follow-up observations by JunoCam onboard the Juno spacecraft carried out 28 hours after the flash 
showed no evident feature at the impact site.
Their non-detection results imply the possible absence of the dark debris seen in
N fragment of the SL9 nucleus \citep{Hammel1995-nn}, 
which is slightly larger than the present impact object.

\begin{figure}
	\includegraphics[width=\columnwidth]{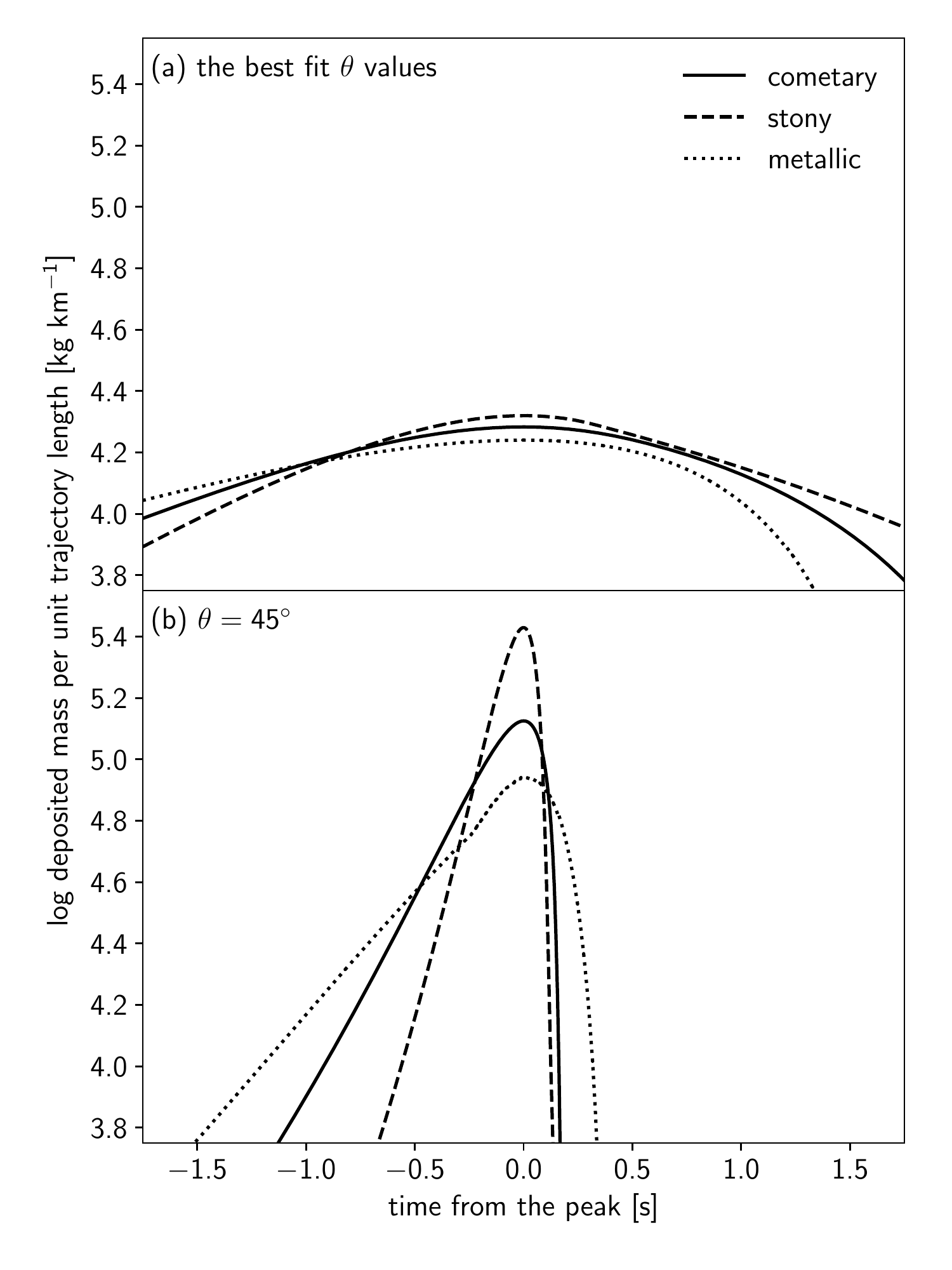}
    \caption{(a) profiles of the deposited mass of the impact object per unit trajectory length derived from the best-fit simulation results as a function of the time from the peak.
    The solid, dashed, and dotted curves represent the profile for the cometary, stony, and metallic cases, respectively.
    (b) the same as (a), but for the models with $\theta = 45\degr$, which is comparable to the angle of entry of the SL9 fragments \citep{Crawford1997-sg}. 
    The other parameters are derived from the best-fit values of the fit.}
    \label{fig:mass}
\end{figure}

\begin{figure}
	\includegraphics[width=\columnwidth]{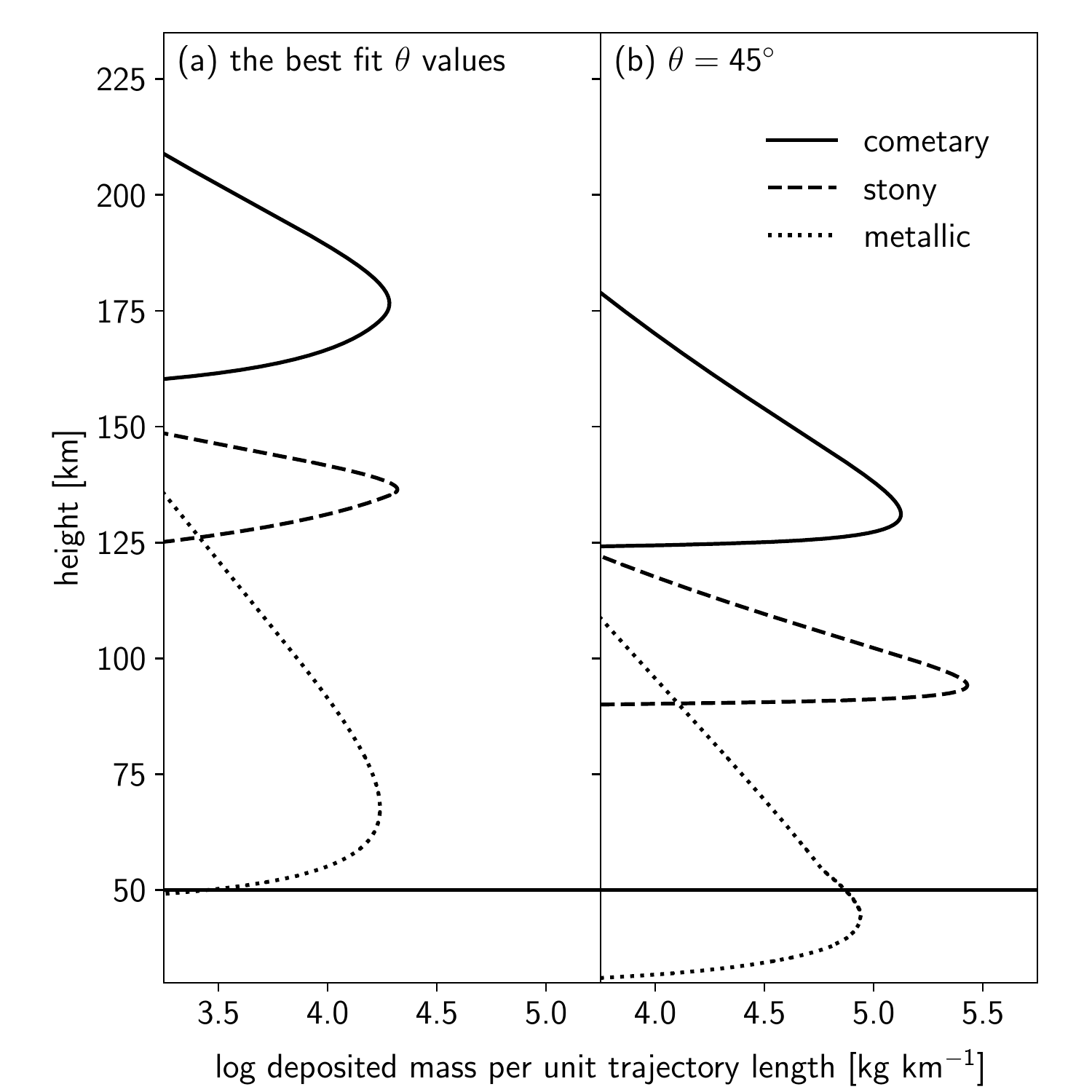}
    \caption{
    (a) profiles of the deposited mass of the impact object per unit trajectory length derived from the best-fit simulation results (the same as Figure~\ref{fig:mass}) as a function of the height $h$.
    The solid, dashed, and dotted curves represent the profile for the cometary, stony, and metallic cases, respectively.
    (b) the same as (a), but for the models with $\theta = 45\degr$, which is comparable to the angle of entry of the SL9 fragments \citep{Crawford1997-sg}. 
    The other parameters are derived from the best-fit values of the fit.
    }
    \label{fig:mass2}
\end{figure}

According to the results of the SL9 studies (e.g., \citealt{Boslough1997-pi}), observable aerosol debris could be created in the dense plume containing evaporated materials from an impact object and Jupiter's entrained atmospheric gas.
Our present results indicate that the angle of entry of the present impact object 
($\leq 12\degr$)
is significantly smaller than those of the SL9 fragments ($\simeq 45\degr$), regardless of the material cases taken into consideration.
Figure~\ref{fig:mass} compares the profiles of the deposited mass of the impact object per unit trajectory length derived from the best-fit simulation results and cases with an angle closer to those of the SL9 impacts.
For all of the three material cases, 
the deposited masses per unit length are approximately an order of magnitude smaller than those for the SL9 cases. 
This is because the mass deposition occurs at higher altitudes, as shown in Figure~\ref{fig:mass2}.
In such high-altitude and low-pressure situations, the impact body suffered lower ablation per unit trajectory than the SL9-like impacts with similar object masses, and the volume densities of the materials would be too small to allow debris particulates to coagulate.

\subsection{Expectations of future megaton-class impacts with different entry conditions}
Since there have been only two events (SL9 and the 2009 event, \citealt{Hammel1995-nn, Sanchez-Lavega2010-rj}) that emerged impact features, 
generation conditions of these features are still unclear.
Detecting and investigating multiple Mt-class impacts with different entry conditions would help to understand the diversities of their consequences,
including the debris generation in Jupiter's atmosphere.
The first detection of an Mt-class impact flash by \citet{Arimatsu2022-if} indicates an occurrence rate of such large-scale impacts on Jupiter is approximately once per year.
Though dedicated monitoring surveys of such infrequent Jovian flashes are thus challenging,
a decade-scale survey dedicated to Jupiter with a sub-meter class telescope(s) would be expected to achieve several detections of the Mt- (and smaller) class impact flashes and their consequences. 

Based on the experiences of the present PONCOTS observations and the encompassing OASES project, 
which is our high-cadence monitoring program of the sky with small telescopes and high-cadence CMOS cameras \citep{Arimatsu2017-lg,Arimatsu2019-zk}, 
we plan to carry out long-term movie monitoring campaign of Jupiter
and the other outer planets
with small dedicated telescopes in the near future.

\section{Conclusions}
\label{sec:con}
The numerical simulations have been carried out to reproduce the observed optical energy profile of 
the PONCOTS flash.
The ablation and fragmentation models with shallow angles of entry approximate the observed optical energy profile of the flash with an extraordinarily long duration. 
The shallower angle of entry possibly resulted in the absence of impact debris features
since the partial pressures of the ablated materials would be too small to allow debris particulates to coagulate. 
The apparent lack of temporal methane absorption change in the observed flash spectrum is also consistent with our best-fit model results. 
The observed optical energy profile is better fitted by the model for cometary or stony cases than that for the metallic case.
Future decadal surveys would make detections of Mt-class impacts and enable us to investigate characteristics of their consequences.

\section*{Acknowledgements}
We thank Erich Karkoschka for providing the data on the methane absorption model, and Liming Li for providing the data on the phase-angle dependence of Jovian albedo. 
This research has been partly supported by JSPS grants (18K13606, 21H01153).

\section*{Data Availability}
The lightcurve data used in this article are available as data behind the article of \citet{Arimatsu2022-if}.
The data products generated in this article are available upon request from the author.



\bibliographystyle{mnras}
\bibliography{ref01} 







\bsp	
\label{lastpage}
\end{document}